\RequirePackage{fix-cm}
\documentclass[smallextended]{svjour3}       
\smartqed  
\usepackage{graphicx}
\usepackage{amssymb,amsmath}
\usepackage{bm,enumerate}
\usepackage{textcomp}
\usepackage[colorlinks=true,citecolor=blue,urlcolor=blue]{hyperref}
\usepackage[utf8x]{inputenc}
\usepackage{cite}

\begin{document}

\title{Two-time correlation functions beyond quantum regression theorem: Effect of external noise}

\titlerunning{Beyond quantum regression theorem}

\author{Arzu~Kurt$^{1}$
}


\institute{Arzu~Kurt\\
\email{arzukurt@ibu.edu.tr}\\
\\
$^1$ Department of Physics, Bolu Abant İzzet Baysal University, Bolu 14030, Turkey.
}

\date{Received: date / Accepted: date}

\maketitle

\begin{abstract}
We present the results of a study of the two-time correlation functions of a dichotomously driven two-level system in contact with a thermal bath by using corrections beyond quantum regression theorem.  In the strong system-environment coupling regime, it is found that the noise parameters could be tuned to control the magnitude of corrections at low environmental temperatures. Motional averaging and narrowing effect of the external noise were observed on the absorption and emission spectra of the two-state system. Furthermore, effects similar to the destruction of tunneling and noise enhancement of transport are observed in the dynamics of the two-time correlations.

\end{abstract}
\section{Introduction}
Developments in spectroscopy techniques have increased the need to develop effective methods to compute the multi-time correlation functions which  encode large amounts of crucial information on the measured spectra~\cite{Scully97,Carmichael99,Gardiner2000,Li2013,Li2018,deVega2017}, that might not be obtained from the expectation values of the  system operators. Use of two-time correlation functions to model measured spectra range from the optical spectrum of the semiconductor quantum dot~\cite{Bertelot2006} and optical emission in quantum-dot-cavity systems~\cite{Winger2009} to Mollow triplet spectra of a resonantly driven quantum dot in a microcavity~\cite{Ulrich2011}.   Jorgensen et. al.'s work on the phonon emission spectrum of the spin-boson model\cite{Jorgensen2020} is a recent  example of modeling the spectral response of open quantum systems by using two-time correlation functions (TTCFs) which has a long history. TTCFs have, also, been used in inferring direct quantum dynamics in experimental investigation of the long-lived electronic quantum coherence for a photosynthetic pigment-protein structure known as Fenna-Matthews-Olsen (FMO) complex~\cite{Brixner2005}. Besides, the authors \cite{Li2013} reported a study of motional averaging and narrowing in the situation where the energy levels of qubit system are modulated by a random telegraph noise in experimental set-up. 

 Many different approaches to calculate the TTCFs have been developed over the years\cite{Onsager31a,Lax63,Alonso2005,deVega2006,Alonso2007,Budini2008,goan11,chen2011,Goan2010,dara2016,deVega2008,Jin2016}. Quantum regression theorem (QRT) provides a direct relation between the dynamics of single-time expectation values of system operators and their two-time correlation functions in Markovian systems and is used extensively in the earlier  modelling attempts~\cite{Onsager31a,Lax63}. 
 Validity of QRT in formulation of TTCFs for non-Markovian systems was questioned by various  groups~\cite{Ford1996,Ford1999,Ford2000,Blanter2000} which lead to various proposals to amend it to account for the memory effects~\cite{Alonso2005,deVega2006,Alonso2007,Budini2008,goan11,chen2011,Goan2010,dara2016,deVega2008,Jin2016}. de Vega and Alonso~\cite{deVega2008} used the stochastic Schr\"{o}dinger equation approach to derive time evolution equations for single and two-time correlation functions of arbitrary system operators and used it to calculate the spectral properties of a two level atom weakly interacting with a highly non-Markovian electromagnetic environment. This model was criticized as being valid only for a limited class of system-environment coupling operators and  the zero temperature environment by Goan et. al.~\cite{goan11} who used a second order master equation approach to systematically improve the QRT evolution equations by including terms that account for the memory effects. They have found considerable differences in time-evolution of the TTCFs obtained in QRT and the improved QRT models even for the weak system-bath coupling regime. Jin et. al. derived  time-local equations of motion for the two-time non-Markovian correlation function of arbitrary system operators in open quantum systems and used them to demonstrate that they could account for the typical non-Markovian effects in current fluctuation spectra of single and double quantum dots.~\cite{Jin2016}. Cosacchi et. al.~\cite{Cosacchi2018} have improved the iterative path integral algorithm for the numerically exact computation of multi-time correlation functions introduced by Shao and Makri~\cite{Shao2002}
 and used it to study spectra of a two-level system simultaneously in contact with Markovian and non-Markovian environments. Their method was found to reproduce the non-Markovian features in the emission of a quantum dot in contact with acoustic phonons. Recent attempts to include multi-time measurement scenario without applying QRT, also exist~\cite{Pollock2018,Jorgensen2019,Jorgensen2020}. Jorgensen and Pollock suggest, for example, an alternative formulation based on time-evolving matrix product operator(TEMPO) algorithm to  both effectively calculate multi-time correlation functions for more general systems and indicate to enhance the efficiency of the method by directly computing the the emission spectra in non-Markovian case for the spin-boson model beyond QRT~\cite{Jorgensen2019}. In order to incorporate multi-time correlation functions Jorgensen and Pollock developed a powerful protocol based on the transfer-tensor framework~\cite{Pollock2018} by examining full of steady-state correlations for the spin-boson model without computing inhomogeneous terms. Also, it is studied the dynamics of system-bath correlations in the spirit of the non-Markovianity relative entropy measurement~\cite{Jorgensen2020}. 

Moreover, various authors suggested the possibility of using the violation of QRT as an indication of non-Markovianity of the system dynamics~\cite{Guarnieri2014,Chen2014,Ali2015,Cosacchi2018}. Guarnieri
 et. al.~\cite{Guarnieri2014} compared the validity of QRT with  CP-divisibility and distinguishability  non-Markovianity measures 
 for the exactly solvable pure dephasing spin boson model and showed that 
 QRT might be violated even when the measures indicate Markovian dynamics. Chen and Manirul Ali have used the non-Markovian TTCFs developed by Goan et. al.~\cite{goan11} to study the violation of Leggett-Garg inequality  for a two-level system in a non-Markovian dephasing environment and reported that the violation decreases with increasing environmental coupling~\cite{Chen2014}. Also, Kurt~\cite{Kurt2020} has considered the violation of QRT for the spin-boson model in the strong coupling regime by using the method developed in Ref.~\cite{goan11} and found that beyond QRT corrections were small for the range of studied parameters.
 
Most of the studies on the non-Markovian two-time correlation functions deals with systems weakly interacting with a non-Markovian thermal environment or Markovian plus another non-Markovian environment. However, some realistic systems, such as FMO light harvesting complexes, are in contact with a thermal environment which might be Markovian or non-Markovian as well as driven by external disturbances which can not be described as a collection of harmonic oscillators~\cite{Jang2018}.
Aim of the present work is to investigate the behaviour of two-time correlation functions for a two-state system (TSS) in a thermal bath when the TSS energy gap is modulated by an external random telegraph signal in the strong coupling regime. Noise averaged time-convolutionless master equation with QRT plus corrections introduced by Goan et. al.~\cite{goan11} is used to examine 
the effect of the external noise on the QRT violations as a function of system, environment and noise parameters.

The paper is organized as the following. We present the model in the polaron frame and the time evolution of a single time as well as two-time correlations functions in non-Markovian case in Sec.~\ref{sec:model}. In Sec.~\ref{sec:results}, we display the result including the averaged over the noise realization for the TTCFs for the non-Markovian system quantities. We briefly summarize the paper in Sec.~\ref{sec:conc.}.

\section{Model}
\label{sec:model}
We consider a two-state system (TSS) that is driven by a random telegraph noise (RTN), which in turn is in contact with a structured bath. The Hamiltonian describing the closed system (TSS + bath) is:
\begin{equation}\label{eq:hamiltonian}
    H=H_{S}(t)+H_B+H_I,
\end{equation}
where $H_S(t)$ is time dependent system Hamiltonian, $H_B$ is the  Hamiltonian of the thermal bath which consists of a set of independent harmonic oscillators, and interaction Hamiltonian between these two is  described by $H_I$. Using the units with $\hbar=1$, the total system Hamiltonian is:
\begin{equation}
H=\frac{\epsilon(t)}{2}\sigma_{z}+
\frac{V}{2}\sigma_{x}+
\sum_{k}\omega_{k}\,b^{\dagger}_{k}\,b_{k}+
\sigma_z\sum_{k} g_{k}\left(b^{\dagger}_{k}+b_{k}\right),
\label{eq:Hamiltonian}
\end{equation}
where $V$ is the tunneling matrix element between the two states,
$\sigma_{i}$ are the Pauli spin matrices,
$b^{\dagger}_{k}(b_{k})$ are creation (annihilation) operators of the $k^{th}$ bath mode,
$g_{k}$ is the interaction strength between the system and the $k$th mode of the environmental oscillators. Here, $\epsilon(t)=\epsilon_0+\Omega\,\alpha(t)$ where $\epsilon_0$ is a bare energy difference between the states of the TSS and the time-dependent term accounts for the random telegraph noise modulation of the state energies. RTN  has two states  with noise amplitude $\pm\Omega$. Its character is described by zero average ($\langle\alpha(t)\rangle=0$) and exponentially decaying auto-correlation function ($\langle\alpha(t)\alpha(t')\rangle=e^{-\nu|t-t'|}$) with noise frequency $\nu$ that is a measure of number of flippings of the noise signal in unit time interval.

In the present study, we consider strong interaction between the TSS and the bath and use polaron transformation to express the interaction  Hamiltonian in a form that lets one use it in perturbative master equation. In the polaron frame, the total time-dependent Hamiltonian of  Eq.~(\ref{eq:hamiltonian}) can be expressed as:
\begin{eqnarray}\label{eq:polaronH}
H^{'}&=& H_{S}^{'}(t)+H_B^{'}+H_I^{'},\nonumber\\
&=&\frac{\epsilon(t)}{2}\sigma_{z}+V_r\,\sigma_x+\sum_{k}\omega_{k}b^{\dagger}_{k}b_{k}+
\sigma_{+}\,B_{-}+\sigma_{-}\,B_{+},
\end{eqnarray}
\noindent where the superscript $"\,'\,"$ means that the operator $O^{'}$ is in the polaron frame. After this point, we will drop the superscript notation, for simplicity. $\sigma_{\pm}$ are the flip operators of the TSS. $B_{\pm}$ are the bath correlation operators and $V_r$ is the reduced tunneling matrix element, and given as a function of system-bath coupling strength $g_{k}$ with the $k^{th}$ oscillator mode respectively:
\begin{eqnarray}
\label{eq:bcf}
B_{\pm}&=&\langle e^{\mp\sum_{k}\frac{2\,g_{k}}{\omega_{k}}\left(b_{k}^{\dagger}-b_{k}\right)}\rangle_{R},\\
V_r&=& V \exp\left(-\frac{1}{4\pi}\int_{0}^{\infty}\frac{J(\omega)}{\omega^2}\coth\left(\beta\omega\right)\right)
\end{eqnarray}
\noindent where $\langle\dots\rangle_R$ implies averaging over the environmental degrees of freedom. It is significant to note that the reduced tunneling rate $V_r$ is equal to zero for the bath spectral densities of which frequency exponent is less than two which is the case in the current study. 

The evolution equation of the reduced density matrix of the TSS in the interaction picture can be written as 
\begin{equation}
\frac{d}{d\,t}\rho_{S}(t)=-\int_{0}^{t}dt_{1}\,\mathrm{Tr}_{R}{\left[H_I(t),\left[H_I(t_{1}),
 \rho_{S}\otimes\rho_{R}\right]\right]}.
\end{equation} 
\noindent where $\mathrm{Tr}_{R}$ indicates partial trace over the environmental modes. In the Schr\"{o}dinger picture, the evolution equations of any system operator $A$ in the second order can be given by

\begin{eqnarray}\label{eq:OneTime}
\frac{d}{d\,t_1}\langle A(t_1)\rangle&=&i\,\mathrm{Tr}_{S\otimes R}\left(\{\left[H_S,A\right]\}(t_1)\,\rho_{T}(0)\right)\nonumber\\
&+&\int_{0}^{t_1}\,d\tau\,\mathrm{Tr}_{S\otimes R}\left(\{\tilde{H}_{I}(\tau-t_1)\left[A,H_{I}\right]\}(t_1)\rho_{T}(0)\right.\nonumber\\
&&\left.+\{\left[H_I,A\right]\tilde{H}_{I}(\tau-t_1)\}(t_1)\rho_{T}(0)\right). 
\end{eqnarray}
\noindent Here, $\tilde{H}_{I}(t)=V\left(\sigma_{-}(t)\,B_{+}(t)+\sigma_{+}(t)\,  B_{-}(t)\right)$ which describes the time evolution in the interaction picture in the polaron frame. Tr with the symbol $S\otimes R$ refers to a trace over the Hilbert space of the total system. Curly brackets in Eqs.~(\ref{eq:OneTime}) and (\ref{eq:TwoTime}) indicate that the expression should be evaluated in the Heisenberg picture and its time should be taken as given in the post bracket.
It is straightforward to express the TTCFs in non-Markovian regime by using the result of Refs.\cite{goan11,dara2016}:
\begin{eqnarray}\label{eq:TwoTime}
\frac{d}{d\,t_1}\langle A(t_1)B(t_2)\rangle&=&i\,\mathrm{Tr}_{S\otimes R}\left(\{\left[H_S,A\right]\}(t_1)B(t_2)\,\rho_{T}(0)\right)\nonumber\\
&+&\int_{0}^{t_1}\,d\tau\,\mathrm{Tr}_{S\otimes R}\left(\{\tilde{H}_{I}(\tau-t_1)\left[A,H_{I}\right]\}(t_1)B(t_2)\rho_{T}(0)\right.\nonumber\\
&&\left.+\{\left[H_I,A\right]\tilde{H}_{I}(\tau-t_1)\}(t_1)B(t_2)\rho_{T}(0)\right)\nonumber\\
&+&\int_{0}^{t_2}d\tau\,\mathrm{Tr}_{S\otimes R}\left(\{\left[H_I,A\right]\}(t_1)\{\left[B,\tilde{H}_{I}(\tau-t_2)\right]\}(t_2)\,\rho_{T}(0)\right).
\end{eqnarray}
\noindent Here, the first two terms on the right hand side of Eq.~(\ref{eq:TwoTime}) belong to the QRT terms while the 
last term accounts for the corrections for the non-Markovian effects. The last integral term in Eq.~(\ref{eq:TwoTime}) is the source of violation of the quantum regression theorem. By using the set of TTCFs in non-Markovian approximation in Eq.(\ref{eq:TwoTime}) we obtain the time evolution of TTCFs of system operators of dichotomously driven spin-boson model in non-Markovian case and average those equations over the large number of noise trajectories in the following section.

\section{Results}
\label{sec:results}

In the present study, we will analyze the dynamics of $\langle\sigma_z(t_1)\sigma_z(t_2)\rangle$ for a TSS which is in contact with a thermal bath at temperature $T$. We will consider the strong  system-bath coupling regime and use a structured spectral density for describing the interaction. The energy levels of the TSS is assumed to be driven stochastically with an RTN signal. By using the Hamiltonian Eq.~(\ref{eq:polaronH}) in the master equation Eq.~(\ref{eq:TwoTime})~\cite{goan11} with $A=\sigma_z$ and $B=\sigma_z$, one can obtain following set of four coupled differential equations for various TTCFs and $\langle\sigma_{z}(t)\rangle$ which include beyond the quantum regression theorem contributions:

\begin{eqnarray}
\frac{d}{d\,t_1}\langle\sigma_{z}(t_1)\,\sigma_{z}(t_2)\rangle&=&-\Gamma_{1}(t_1)\langle\sigma_{z}(t_1)\sigma_{z}(t_2)\rangle-\Gamma_2(t_1)\langle\sigma_z(t_2)\rangle\nonumber\\
&&+4\,\Gamma_{3}(t_1,t_2)\langle\sigma_{-}(t_1)\sigma_{+}(t_2)\rangle+4\,\Gamma_{4}(t_1,t_2)\langle\sigma_{+}(t_1)\sigma_{-} (t_2)\rangle\label{eq:szz}\\
\frac{d}{d\,t_1}\langle\sigma_{z}(t_1)\rangle&=&-\Gamma_1(t_1)\langle\sigma_{z}(t_1)\rangle-\Gamma_{2}(t_1)\label{eq:sz},\\
\frac{d}{d\,t_1}\langle\sigma_{+}(t_1)\sigma_{-}(t_2)\rangle&=&\left[i\,\left(\epsilon_{0}+\Omega\,\alpha(t_1)\right)-\Gamma_{5}(t_1)\right]\langle\sigma_{+}(t_1)\sigma_{-}(t_2)\rangle\nonumber\\
&&+\Gamma_{3}(t_1,t_2)\langle\sigma_{z}(t_1)\sigma_{z}(t_2)\rangle
\label{eq:spm},\\
\frac{d}{d\,t_1}\langle\sigma_{-}(t_1)\sigma_{+}(t_2)\rangle&=&-\left[i\,\left(\epsilon_{0}+\Omega\,\alpha(t_1)\right)+\Gamma_{5}(t_1)^{*}\right]\langle\sigma_{-}(t_1)\sigma_{+}(t_2)\rangle\nonumber\\
&&+\Gamma_{4}(t_1,t_2)\langle\sigma_{z}(t_1)\sigma_{z}(t_2)\rangle
\label{eq:smp},
\end{eqnarray}
\noindent We should note that derivation of Eqs.~(\ref{eq:szz})-(\ref{eq:smp}) is simplified by the fact that the reduced tunneling 
rate $V_r$ in Eq.~(\ref{eq:polaronH}) equals to zero for the chosen spectral density. All the terms in Eqs.~(\ref{eq:szz})-(\ref{eq:smp}) that contain $\Gamma_i$ coefficients with $i=3$ and 4 are due to the corrections to the QRT (the last line of Eq.~(\ref{eq:TwoTime})) while those with $i=1,2,5$ and $6$ account for the quantum regression theorem. The environmental correlation functions $\Gamma_i(t)$ are derived from thermal averages of various system and bath operators as: 

 \begin{eqnarray*}
\Gamma_{1}(t_1)&=&4\,V^{2}\int_{0}^{t_1}d\tau\,e^{-Q_2\left(t_1-\tau\right)}\cos{\left[Q_{1}(t_1-\tau)\right]}\,\cos{\left[ f(t_1,\tau)\right]},\\
\Gamma_{2}(t_1)&=&4\,V^{2}\int_{0}^{t_1}d\tau\,e^{-Q_{2}\left(t_1-\tau\right)}\,\sin{\left[Q_{1}(t_1-\tau)\right]}\,\sin{\left[ f(t_1,\tau)\right]},\\
\Gamma_{3}(t_1,t_2)&=&V^{2}\int_{0}^{t_2}d\tau\,e^{-Q_2\left(t_1-\tau\right)+i\,Q_{1}\left(t_1-\tau\right)}\,e^{i\,f(t_2,\tau)},\\
\Gamma_{4}(t_1,t_2)&=&V^{2}\int_{0}^{t_2}d\tau\,e^{-Q_2\left(t_1-\tau\right)+i\,Q_{1}\left(t_1-\tau\right)}\,e^{-i\,f(t_2,\tau)},\\
\Gamma_{5}(t_1)&=&2\,V^{2}\int_{0}^{t_1}d\tau\,e^{-Q_{2}\left(t_1-\tau\right)}\,\cos{\left[Q_{1}(t_1-\tau)\right]}\,e^{i\,f(t_1,\tau)},\\
\end{eqnarray*}
\noindent where
\begin{eqnarray}
Q_{1}(t)&=&\frac{1}{2\,\pi}\int_{0}^{\infty}d\omega\frac{J(\omega)}{\omega^{2}}\sin(\omega t),\\
Q_{2}(t)&=&\frac{1}{2\,\pi}\int_{0}^{\infty}d\omega\frac{J(\omega)}{\omega^{2}}\coth\left(\frac{\beta\,\omega}{2}\right)
(1-\cos(\omega t)),\\
f(t,\tau)&=&\epsilon_{0}(t-\tau)+\Omega\,\int_{t}^{\tau}d\zeta\,\alpha(\zeta)\label{eq:int_RTN},
\end{eqnarray}
\noindent where $Q_{1,2}(t)$ are the imaginary and  the real parts of the bath correlation function. $f(t,\tau)$ accounts for the static as well as stochastically fluctuating system energy gap. Spectral density $J(\omega)$ characterizes the interaction between the TSS and the environment as a function of environmental modes. We choose $J(\omega)$ as the structured spectral density. In this model, the environment is described by single harmonic oscillator (with frequency $\omega_0$) which is in contact with an Ohmic thermal bath that broadens its energy levels~\cite{Garg1985}:

\begin{equation}
\label{eq:spectral}
    J(\omega)=8\,\kappa^{2}\frac{\gamma\,\omega_0\,\omega}{(\omega^{2}-\omega_0^2)^2+4\gamma^2\,\omega^2},
\end{equation}
\noindent where $\kappa$ is the magnitude of system-bath coupling, $\omega_0$ is the frequency of the central harmonic oscillator, $\gamma$ is the broadening of the oscillator levels due to interaction with the environment.  A rough measure of the strength of the interaction between the TSS and its environment is the reorganization energy of the bath  ($E_r=\int_{0}^{\infty}d\omega\,J(\omega)/\omega$). The reorganization energy for the chosen $J(\omega)$ is equal to $\kappa^2/\omega_0$.

\subsection{Stochastic Averaging}

The RTN signal $\alpha(t)$ appears in the coupled differential equations that describe the dynamics of the two-time correlation functions in Eqs.(\ref{eq:szz})-(\ref{eq:smp}) as bare coefficients as well as its integral in definition of $\Gamma_i(t)$ coefficients. One can either use an ensemble averaging approach to find the noise averaged TTCFs by solving  Eqs.~(\ref{eq:szz})-(\ref{eq:smp}) for an appropriately controlled number of different realizations of RTN noise and average the solutions or use probability density of RTN signal to average the coupled differential equations and solve those averaged equations. Thanks to theorems due to Bourret-Frisch~\cite{Bourret73} which provide a  factorization for the multi-time correlation functions of dichotomous process and Loginov-Shapiro~\cite{Shapiro78} which relates the average of time evolution operator of the noise with that of the system operators,  Eqs.~(\ref{eq:szz})-(\ref{eq:smp}) can be averaged exactly over the RTN~\cite{Magazzu2017}. The averaging process effectively doubles the number of correlation functions by coupling the correlations of the noise signal $\alpha(t)$ with the system operator correlations.  We will adopt the single average notation to describe both the correlation and noise averaging, i. e., $\langle\sigma_{z}(t)\rangle$ denotes noise averaged correlation of $\sigma_z$ at times $t_1$ and $t_2$ in the rest of the paper. The set of RTN averaged coupled differential equations for $\langle\sigma_z(t_1)\rangle$ and $\langle\alpha(t_1)\sigma_z(t_2)\rangle$ are:
\begin{eqnarray}
\frac{d}{d\,t_1}\langle\sigma_{z}(t_1)\rangle&=&-\Gamma_1(t_1)\langle\sigma_z(t_1)\rangle
                 +\Gamma_{2}(t_1)\langle\alpha(t_1)\,\sigma_{z}(t_1)\rangle
                 -\Gamma_{3}(t_1),\nonumber\\
\frac{d}{d\,t_1}\langle\alpha(t_1)\sigma_{z}(t_1)\rangle&=&-(\nu+\Gamma_1(t_1))\,\langle\alpha(t_1)\sigma_{z}(t_1)
            \rangle+\Gamma_{2}(t_1)\langle\sigma_{z}(t_1)\rangle
                   -\Gamma_{4}(t_1).
                   \label{eq:av_sz}
\end{eqnarray}

The noise averaged coupled differential equations for the six two-time correlation functions can be expressed as:
\begin{equation}
    \frac{d}{dt_1}Y=A(t_1,t_2)\cdot Y+b(t_1,t_2)
    \label{eq:avMatrix}
\end{equation}
\noindent where $Y$ is a column vector whose elements are the different correlations:
\begin{eqnarray*}
    Y=\left(
    \langle\sigma_{z}(t_1)\sigma_{z}(t_2)\rangle,
    \langle\alpha(t_1)\sigma_{z}(t_1)\sigma_{z}(t_2)\rangle,
    \langle\sigma_{+}(t_1)\sigma_{-}(t_2)\rangle,
    \langle\alpha(t_1)\sigma_{+}(t_1)\sigma_{-}(t_2)\rangle,\right.\\
    \left.
    \langle\sigma_{-}(t_1)\sigma_{+}(t_2)\rangle,
    \langle\alpha(t_1)\sigma_{-}(t_1)\sigma_{+}(t_2)\rangle
  \right)^T
\end{eqnarray*}
\noindent and $A$ is the time-dependent coefficient matrix of the noise-averaged coupled differential equations:
\begin{eqnarray*}
A=\left(\begin{array}{cccccc}
-\Gamma_{1,1}&\Gamma_{1,2}&-4\Gamma_{3,1}&-4\Gamma_{3,2}&4\Gamma_{4,1}&4\Gamma_{4,2}\\
\Gamma_{1,2}&-\left(\nu+\Gamma_{1,1}\right)&4\Gamma_{3,2}&4\Gamma_{3,1}&-4\Gamma_{4,2}&4\Gamma_{4,1}\\
\Gamma_{4,1}&-\Gamma_{4,2}&(i\epsilon_0-\Gamma_{5,1})&(i\Omega+\Gamma_{5,2})&0&0\\
-\Gamma_{4,2}&\Gamma_{4,1}&(i\Omega+\Gamma_{5,2})&-(\nu-i\epsilon_0+\Gamma_{5,1})&0&0\\
\Gamma_{3,1}&\Gamma_{3,2}&0&0&-(i\epsilon_0+\Gamma_{6,1})&-(i\Omega+\Gamma_{6,2})\\
\Gamma_{3,2}&\Gamma_{3,1}&0&0&-(i\Omega+\Gamma_{6,2})&-(\nu+i\epsilon_0+\Gamma_{6,1})
\end{array}
\right)
\end{eqnarray*}
\noindent where all $\Gamma$s in $A$ depend on time $t_1$ except $\Gamma_{3,}$ and $\Gamma_{4,}$ which have double time arguments $(t_1,t_2)$. The $b(t_1,t_2)$ term in Eq.~(\ref{eq:avMatrix}) contributes to the change in   $\langle\sigma_{z}(t_1)\sigma_{z}(t_2)\rangle$ and $\langle\alpha(t_1)\sigma_{z}(t_1)\sigma_{z}(t_2)\rangle$ terms and is found to be:
\begin{eqnarray*}
b=\left(-\Gamma_{2,1}(t) g_1(t_2)-e^{-\nu|t-t_2|}\Gamma_{2,2}(t) g_2(t_2),-\Gamma_{2,2}(t) g_1(t_2)-e^{-\nu|t-t_2|}\Gamma_{2,1}(t)g_2(t_2),0,0,0,0\right)^{T}
\end{eqnarray*}
\noindent where $g_1(t)=\langle\sigma_{z}(t)\rangle$ and $g_2(t)=\langle \alpha(t)\sigma_{z}(t)\rangle$. $\Gamma_{i,j}$s of $A$ involve the time evolution operator $S(t,t')=\exp\left[-i\,\Omega\int_{t'}^{t}d\tau\alpha(\tau)\right]$ of the Kubo oscillator and 
are given as:

\begin{eqnarray}
\label{eq:kernels-6-6}
\Gamma_{1,1}(t)&=&\int_{0}^t d\tau\,\mathcal{E}_{cc}(t-\tau)S_{0}(t-\tau)\nonumber\\
\Gamma_{1,2}(t)&=&i\int_{0}^t d\tau\,\mathcal{E}_{cs}(t-\tau)S_{1}(t-\tau)\nonumber\\
\Gamma_{2,1}(t)&=&\int_{0}^t d\tau\,\mathcal{E}_{ss}(t-\tau)S_{0}(t-\tau)\nonumber\\
\Gamma_{2,2}(t)&=&i\int_{0}^t d\tau\,\mathcal{E}_{sc}(t-\tau)S_{1}(t-\tau)\nonumber\\
\Gamma_{3,1}(t,t_2)&=&\int_{0}^{t_2} d\tau\,\mathcal{E}_{f+}(t-\tau)e^{i\epsilon(t_2-\tau)}S_{0}(t_2-\tau)\nonumber\\
\Gamma_{3,2}(t,t_2)&=&\int_{0}^{t_2} d\tau\,\mathcal{E}_{f+}(t-\tau)e^{i\epsilon(t_2-\tau)}S_{1}(t_2-\tau)\\
\Gamma_{4,1}(t,t_2)&=&\int_{0}^{t_2} d\tau\,\mathcal{E}_{f-}(t-\tau)e^{-i\epsilon(t_2-\tau)}S_{0}(t_2-\tau)\nonumber\\
\Gamma_{4,2}(t,t_2)&=&\int_{0}^{t_2} d\tau\,\mathcal{E}_{f-}(t-\tau)e^{-i\epsilon(t_2-\tau)}S_{1}(t_2-\tau)\nonumber\\
\Gamma_{5,1}(t)&=&\int_{0}^t d\tau\,\mathcal{E}_{c-}(t-\tau)S_{0}(t-\tau)\nonumber\\
\Gamma_{5,2}(t)&=&\int_{0}^t d\tau\,\mathcal{E}_{c-}(t-\tau)S_{1}(t-\tau)\nonumber\\
\Gamma_{6,1}(t)&=&\int_{0}^t d\tau\,\mathcal{E}_{c+}(t-\tau)S_{0}(t-\tau)\nonumber\\
\Gamma_{6,2}(t)&=&\int_{0}^t d\tau\,\mathcal{E}_{c+}(t-\tau)S_{1}(t-\tau)\nonumber
\end{eqnarray}
\noindent where $\mathcal{E}_{cs}(t)=e^{-Q_{2}(t)}\mathrm{(c)os}\left(Q_{1}(t)\right)\,\mathrm{(s)in}\left(\epsilon_{0}\,t\right)$, $\mathcal{E}_{c\pm}(t)=e^{-Q_{2}(t)}\mathrm{cos}\left(Q_{1}(t)\right)\,e^{\pm\,i\,\epsilon_{0}\,t}$, and $\mathcal{E}_{f\pm}(t)=e^{-Q_{2}(t)}e^{i(Q_{1}(t)\pm\epsilon_0\,t)}$. $S_{0}(t)$ and $S_{1}(t)$ in kernel definitions are, also, dichotomous noise propagators~\cite{Goychuk95} which can be expressed as:
\begin{eqnarray}
S_{0}(t)&=&\frac{1}{2\eta}\left(\nu_{+}e^{-t\,\nu_{-}/2}-\nu_{-}e^{-t\,\nu_{+}/2}\right),\nonumber\\
S_{1}(t)&=&i\frac{\Omega}{\nu}\left(e^{-t\,\nu_{+}/2}-e^{-t\,\nu_{-}/2}\right)
\end{eqnarray}
\noindent where $\eta=\sqrt{\nu^{2}-4\Omega^{2}}$, $\nu_{+}=\nu+\eta$, and $\nu_{-}=\nu-\eta$.

 One should note that the $\Gamma_{i,j}$ that depends on a single time argument 
stem from the non-Markovian quantum regression theorem, while the ones with double time argument arise as corrections to the QRT. They will be referred to as QRT$+$ in the rest of the current paper. Also, when there is no external noise, $S_0(t)=1$ and $S_1(t)=0$, so all the $\Gamma_{i,2}$ are zero.

\begin{figure}
\begin{center}
     \begin{tabular}[b]{c}
    \includegraphics[width=0.4\linewidth]{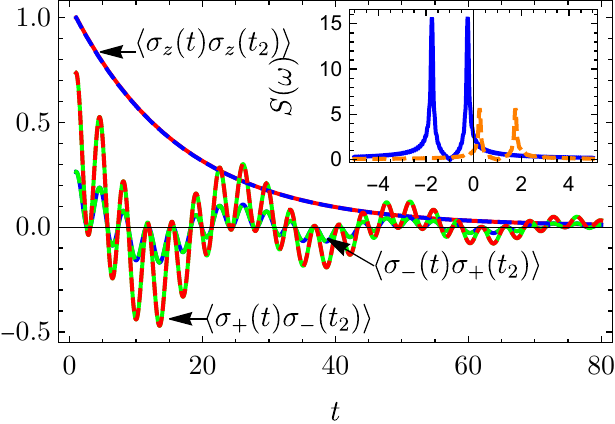}
   \\ \small (a) {$\epsilon_0=1,\,\beta=0.02,\,\nu=0.01$}
    \end{tabular}\qquad
    \begin{tabular}[b]{c}
    \includegraphics[width=0.4\linewidth]{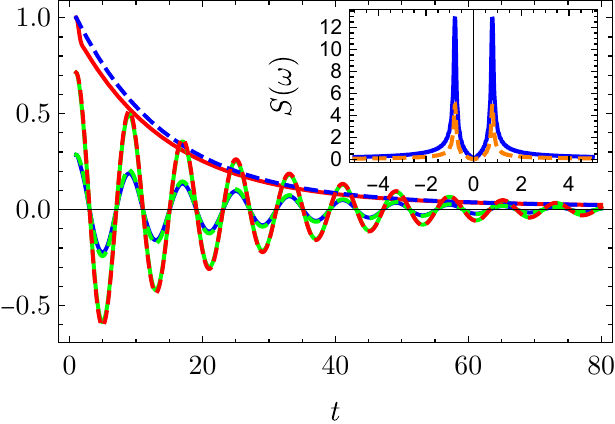}
   \\ \small (b) {$\epsilon_0=0,\,\beta=50,\,\nu=0.01$}
    \end{tabular}\qquad
    \begin{tabular}[b]{c}
    \includegraphics[width=0.4\linewidth]{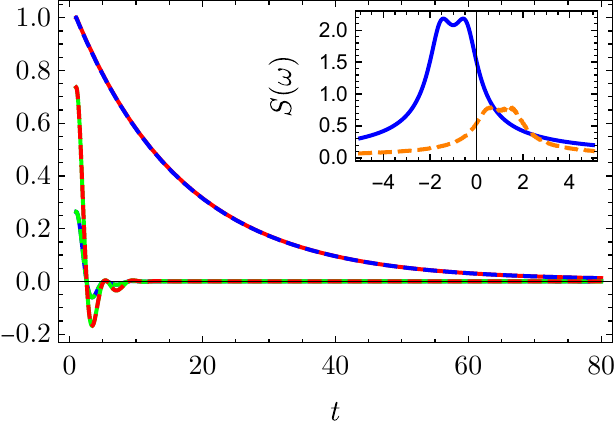}
   \\ \small (c) {$\epsilon_0=1,\,\beta=0.02,\,\nu=1$}
    \end{tabular}\qquad
    \begin{tabular}[b]{c}
    \includegraphics[width=0.4\linewidth]{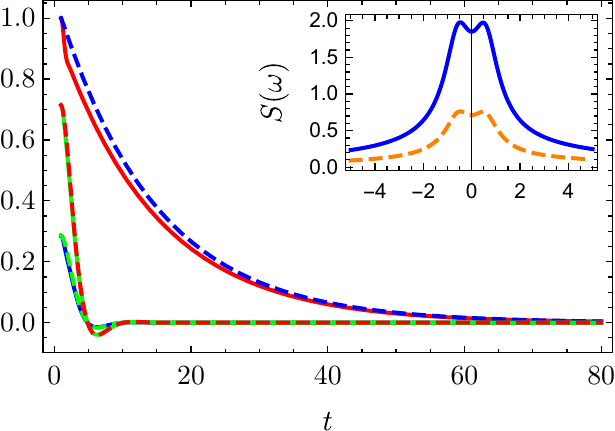}
   \\ \small (d) {$\epsilon_0=0,\,\beta=50,\,\nu=1$}
    \end{tabular}\qquad
    \end{center}
    \caption{Dynamics of the real part of two-time correlations $\langle\sigma_{z}(t)\sigma_z(t_2)\rangle$, $\langle\sigma_{+}(t)\sigma_{-}(t_2)\rangle$ and $\langle\sigma_{-}(t)\sigma_{+}(t_2)\rangle$ for the QRT (dashed line) and QRT+ (solid line) approaches for various combinations of the TSS bias, the noise frequency and  temperature of the environment. Inset shows the power spectra of  $\langle\sigma_{+}(t)\sigma_{-}(t_2)\rangle$ (solid line) and $\langle\sigma_{+}(-)\sigma_{+}(t_2)\rangle$ (dashed line) correlations calculated with the QRT+ corrections. The noise amplitude $\Omega=0.75$ for all the figures.}

\label{fig:dynFT}
\end{figure}

For the present study, we consider the strong system-environment coupling regime with 
$\omega_0=1$, $\kappa=2$ along with a relatively large dispersion $\gamma=0.5$ and investigate the low ($\beta=50$) and the high temperature ($\beta=0.02$) as well as the biased ($\epsilon_{0}=1$) and the non-biased ($\epsilon_{0}=1$) cases. With these coupling parameters, the short time approximation\cite{Garg1985} can be used to write the bath correlation 
functions in a simpler form as $Q_{1}(t)=E_r t$ and $Q_{2}(t)=-\xi t^2$ where 
\begin{equation*}
    \xi=E_r/\beta+\kappa^2/\left(\pi \sqrt{\omega_0^2-\gamma^2}\right)\omega_0\,
\mathrm{Im}\left[\Psi\left(0,1+\beta\left(\gamma+i\sqrt{\omega_0^2-\gamma^2}\right)/(2\pi)\right)\right].
\end{equation*}
\noindent For this choice of the bath correlation functions, the $\Gamma_{i,j}$ coefficients in Eq.~(\ref{eq:kernels-6-6}) can be expressed as sum of error functions of various sum and difference combinations of the TSS bias, noise frequency and the bath reorganization energy. But, the expressions are too cumbersome and will not be displayed here.

We start the discussion of the results with presentation of the various qualitatively different forms of time-dependence of the two-time correlation functions calculated in non-Markovian QRT and QRT+ approximations in Fig.~\ref{fig:dynFT}a-d. While $\langle\sigma_z(t_{1})\sigma_z(t_{2})\rangle$ is found to decay exponentially with a rate that depends on the temperature, the bias and the properties of the noise for the strong system-bath coupling, $\langle\sigma_{+}(t_{1})\sigma_{-}(t_2)\rangle$ and $\langle\sigma_{-}(t_1)\sigma_{+}(t_2)\rangle$ display richer dynamics. It can be seen from the figures that the difference between the QRT and QRT+ approximations for all three correlation functions at low temperature is higher than the  difference at high temperature. The difference between the two will be discussed in more detail below. 

\begin{figure}
\begin{center}
    \begin{tabular}[b]{c}
    \includegraphics[width=0.4\linewidth]{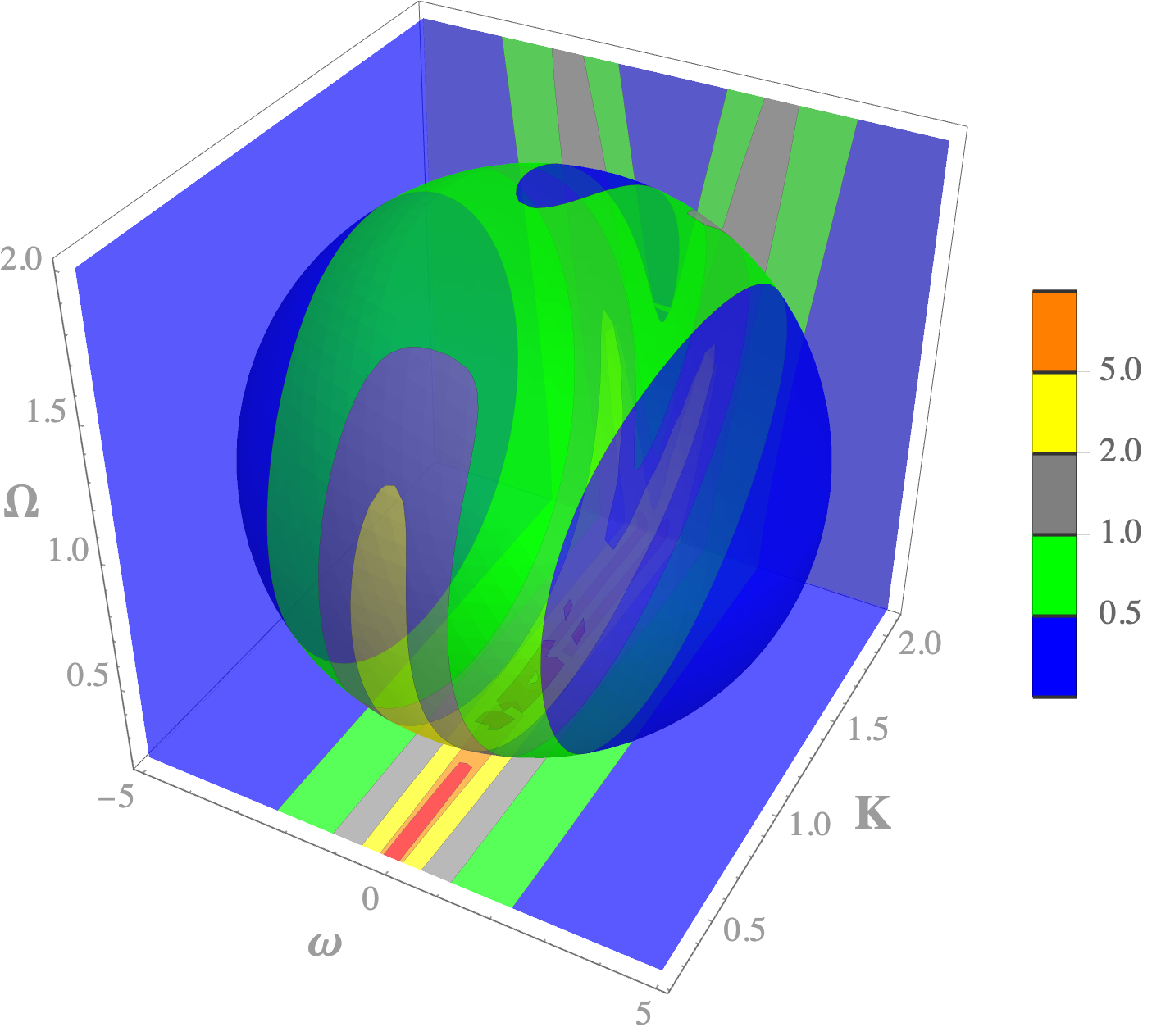}
   \\ \small (a) {$\beta=50$}
    \end{tabular}\qquad
    \begin{tabular}[b]{c}
    \includegraphics[width=0.4\linewidth]{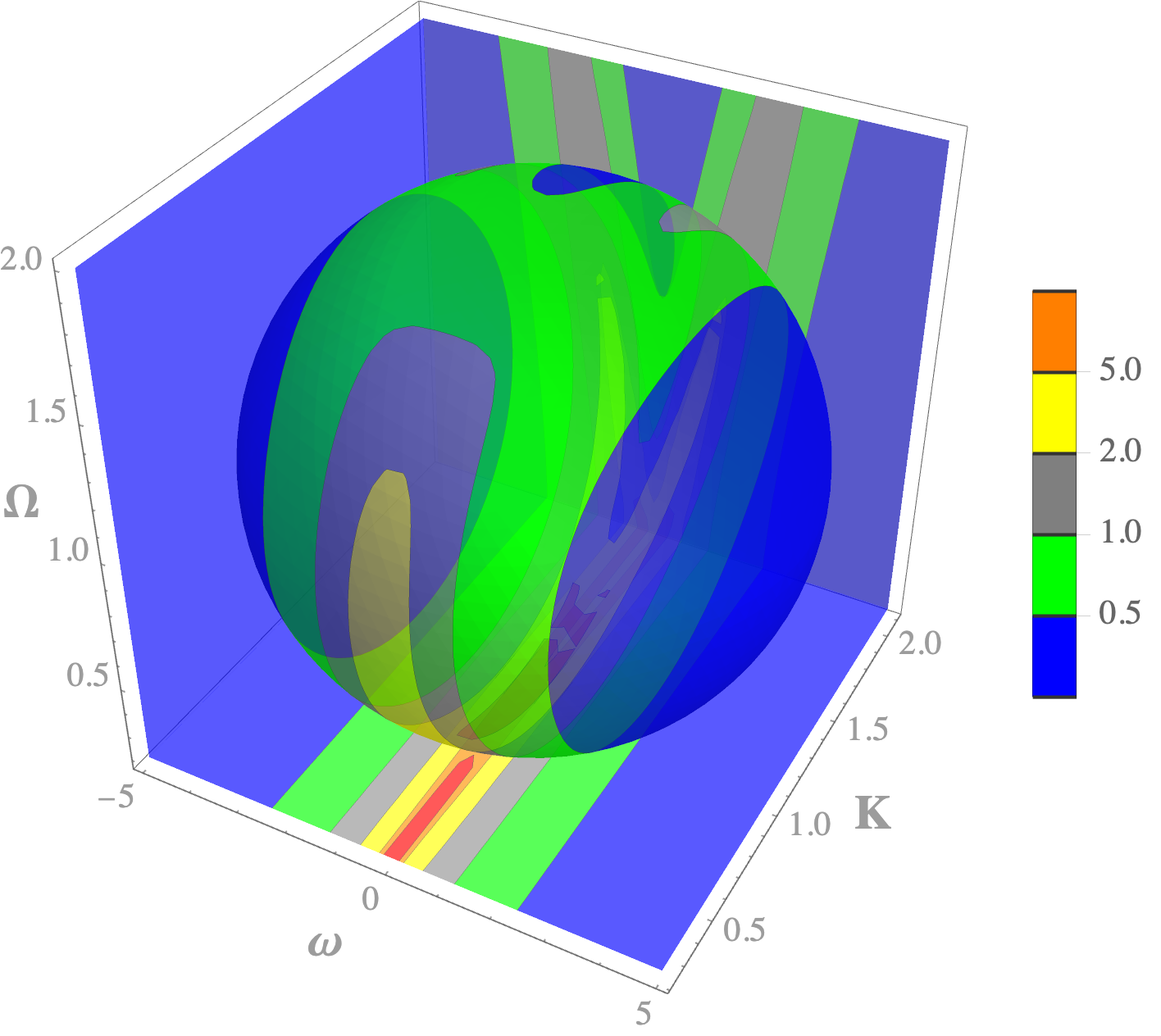}
    \\\small (b) {$\beta=0.02$}
    \end{tabular}\qquad
    \end{center}
    \caption{Noise amplitude and color dependence of absorption spectra at low and high temperature for the non-biased TSS.}
    \label{fig:plot4D}
\end{figure}

The dynamics of a two state system driven by RTN can be deduced intuitively at the asymptotic limits with the help of the concept of the noise color or the Kubo number $K=\Omega/\nu$. As the noise jumps of magnitude $\Delta E=\hbar\Omega$ occur with average time spacing $\Delta t=1/\nu$, the energy-time uncertainty relation ($\Delta E\Delta t=\hbar\Omega/\nu\ge \hbar/2$) would make resolving those fluctuations impossible at the fast jumping limit. Resulting motional averaging leads to emission or absorption at dynamically averaged $\epsilon_0$ frequency although the system is at one of the states $\epsilon_0\pm \Omega$ at all times. At the slow noise limit, the mean time between the jump events is long and the dynamics of the system can be considered as the static average at frequencies $\epsilon_0\pm\Omega$. Fourier transform of $\langle\sigma_{-}(t+\tau)\sigma_{+}(t)\rangle$ and $\langle\sigma_{+}(t+\tau)\sigma_{-}(t)\rangle$ are measures of absorption and emission spectrum, respectively. The insets in Fig.~\ref{fig:dynFT}a-d display the power spectrum of the correlations $\langle\sigma_{+}(t_{1})\sigma_{-}(t_2)\rangle$ and $\langle\sigma_{-}(t_{1})\sigma_{+}(t_2)\rangle$. The first row where $K=75$ is the slow noise (strongly colored noise) limit and one can observe emission and absorption phenomena at  frequencies $\epsilon_0\pm\Omega$ clearly. The plots in the second row display the dynamics of the correlation at intermediate noise color with $K=0.75$ for which the motional averaging starts to dominate the power spectrum as the displayed spectra tend to broaden and become a single peak at frequency $\epsilon_0$. We display the $\Omega$, $K$ and frequency dependence of the absorption spectrum at high and low temperatures in Fig.\ref{fig:plot4D}. As can be seen from the plots, the single peak to double peak behaviour start emerging around $K\approx 0.7$ independent of the bath temperature.
\begin{figure}
\begin{center}
    \begin{tabular}[b]{c}
    \includegraphics[width=0.4\linewidth]{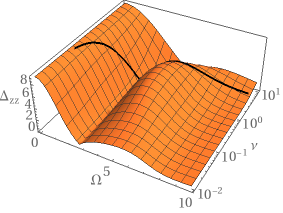}
    \\\small (a) {$\langle\sigma_{z}(t_1)\sigma_{z}(t_2)\rangle$}
    \end{tabular}\qquad
    \begin{tabular}[b]{c}
    \includegraphics[width=0.4\linewidth]{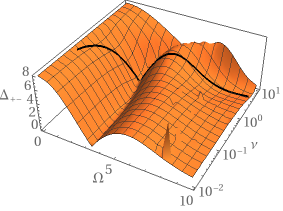}
    \\\small (b) {$\langle\sigma_{-}(t_1)\sigma_{+}(t_2)\rangle$}
    \end{tabular}\qquad
    \end{center}
    \caption{Noise frequency $\nu$ and amplitude $\Omega$ dependence of the relative difference between the QRT and QRT+ ($\epsilon_{0}=0,\,\beta=50$). Thick line indicates $\nu=2\Omega$.}
    \label{fig:difference}
\end{figure}

There has been some discussion on the relevance of QRT$+$ terms on the non-Markovianity of the dynamics of the system, which suggests that the difference can be used as a measure of non-Markovianity~\cite{Ali2015,Cosacchi2018,Guarnieri2014,Chen2014}. As a rough measure of the difference between the two dynamics, we adopt a heuristic approach and define
\begin{equation}\label{eq:measure}
    \Delta_{c}=100 \left|\frac{I_{\textrm{QRT}}(c)-I_{\textrm{QRT+}}(c)}{I_{\textrm{QRT}}(c)}\right|,
\end{equation}
\noindent where $I_{QRT}(c)$ is the integral of the absolute value of the correlation function $c$ obtained as solution of the QRT formulation. $\Delta_{c}$ is found to be highly sensitive to the environmental temperature for the system and the other environment parameters considered in the present study. It is found to be close to zero ($<0.25$) at high temperature ($\beta=0.02$) for all three correlation functions. Figure~\ref{fig:difference} displays $\Delta_{zz}$ and $\Delta_{-+}$ as function of the noise frequency and amplitude at low temperature ($\beta=50$) for the non-biased system. $\Delta_{zz}$ and $\Delta_{-+}$ are found to show similar qualitative and quantitative dependence on noise parameters. Interestingly, for both correlations, the difference display a death and revival cycle as function of the noise amplitude. The cycle has a weak dependence on the noise frequency which is peculiar because one would expect effect of noise to be dependent on its color or Kubo number.

\begin{figure}
\begin{center}
     \begin{tabular}[b]{c}
    \includegraphics[width=0.4\linewidth]{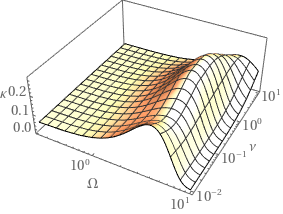}
   \\ \small (a) {$\epsilon_{0}=0,\,\beta=50$}
    \end{tabular}\qquad
    \begin{tabular}[b]{c}
    \includegraphics[width=0.4\linewidth]{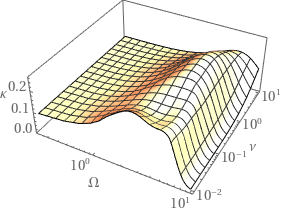}
  \\\small (b) {$\epsilon_{0}=1,\,\beta=50$}
    \end{tabular}\qquad
    \begin{tabular}[b]{c}
    \includegraphics[width=0.4\linewidth]{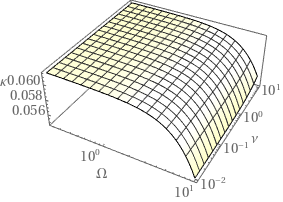}
    \\ \small (c) {$\epsilon_{0}=0,\,\beta=0.02$}
    \end{tabular}\qquad
    \begin{tabular}[b]{c}
    \includegraphics[width=0.4\linewidth]{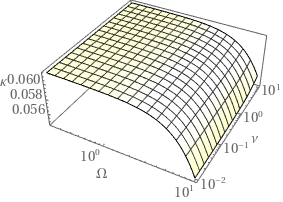}
   \\ \small (d) {$\epsilon_{0}=1,\,\beta=0.02$}
    \end{tabular}\qquad
    \end{center}
    \caption{Noise frequency $\nu$ and amplitude $\Omega$ dependence of the inverse life time of the $\langle\sigma_{z}(t_1)\sigma_{z}(t_2)\rangle$.}
    \label{fig:ks}
   \end{figure}

The two-time correlation function $\langle\sigma_{z}(t_1)\sigma_{z}(t_2)\rangle$ decays exponentially for the system and environment parameters chosen for the present study as can be seen from Figs.~\ref{fig:dynFT}a-d. We fit the time dependence of $\langle\sigma_{z}(t_1)\sigma_{z}(t_2)\rangle$ to a function of the form $p+(1-p)e^{-kt}$ where $p$ is the limit of $\langle\sigma_{z}(t_1)\sigma_{z}(1)\rangle$ as $t_1 \to \infty$ and $k$ is the decay parameter of the correlation. We display the noise frequency and amplitude dependence of this exponential decay parameter $k$ at different temperature and TSS bias parameters in Fig.~\ref{fig:ks} a-d. It can be seen from the sub-figures that $k$ decreases towards zero at high noise amplitude irrespective of the bath temperature, TSS bias and the noise frequency which is reminiscent of the noise induced destruction of tunneling~\cite{Berry95}. Effect of $\nu$ and $\Omega$ on $k$ is found to be observably different at low and high temperatures. At high temperature, $k$ is almost independent of the bias and the noise frequency and it decreases with increasing noise amplitude monotonously while one can observe an enhancement of the transport especially for the low-frequency noise for both the biased and non-biased TSS at low temperature. This finding is another example of so called ENAQT~\cite{Rebentrost2009} which describes noise assisted enhancement of quantum transport.

\begin{figure}
\begin{center}
    \begin{tabular}[b]{c}
    \includegraphics[width=0.4\linewidth]{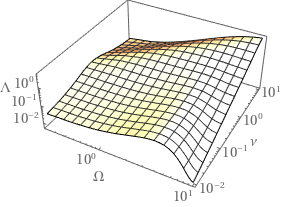}
    \\ \small (a) $\epsilon_{0}=0,\,\beta=50$
    \end{tabular}\qquad
    \begin{tabular}[b]{c}
    \includegraphics[width=0.4\linewidth]{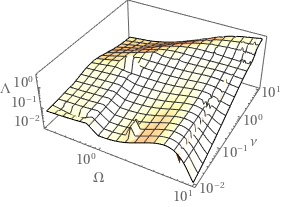}
    \\ \small (b) $\epsilon_{0}=1,\,\beta=50$
    \end{tabular}\qquad
    \begin{tabular}[b]{c}
    \includegraphics[width=0.4\linewidth]{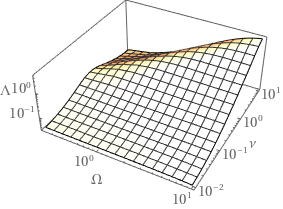}
   \\ \small (c){ $\epsilon_{0}=0,\,\beta=0.02$}
     \end{tabular}\qquad
     \begin{tabular}[b]{c}
    \includegraphics[width=0.4\linewidth]{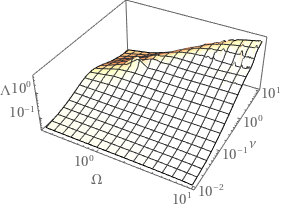}
   \\ \small (d) $\epsilon_{0}=1,\,\beta=0.02$
     \end{tabular}\qquad
     \end{center}
    \caption{Noise frequency $\nu$ and amplitude $\Omega$ dependence of the decoherence rate for  $\langle\sigma_{-}(t_1)\sigma_{+}(t_2)\rangle$.}
    \label{fig:t2}
\end{figure}

We found that the real part of the correlation $\langle\sigma_{-}(t_1)\sigma_{+}(t_2)\rangle$ could be fitted to a function of the form $e^{-\Lambda t} \left(a_1\cos{\left(\omega_1 t\right)}+a_2\cos{\left(\omega_2 t\right)}\right)$ which could in return be used to extract the exponential decay rate $\Lambda$ which is similar to the decoherence rate of the $\langle\sigma_x\rangle$. Figure~\ref{fig:t2} indicates the noise amplitude and frequency dependence of $\Lambda$ at high and low temperatures for the biased and non-biased TSS. The $\Lambda$ value at low $\Omega$ might be considered as the $1/T_2$~\cite{Wang2005} of the system. As expected, its value at high temperature is higher than at low temperature. Decoherence rate $\Lambda$ is found to be high at large $\nu$ and $\Omega$ values irrespective of the environmental temperature and the bias of the TSS which is expected because dynamics is dominated by the noise. An interesting decrease of decoherence rate can be observed at low temperatures for the slow noise for both the biased and non-biased TSS (Fig.~\ref{fig:t2}a and b). Also, for the very weak noise strength, the decoherence rate displays a resonance structure as function of the noise frequency as can be seen in all four sub-plots in Fig.~\ref{fig:t2}.

\section{Conclusions}
\label{sec:conc.}
We have investigated the effect of random telegraph noise on the two-time correlation functions of a strongly coupled spin-boson model with structured spectral density. Both the quantum regression theorem and corrections to the QRT terms were included in the formulation of the master equations for the correlations. One of the important findings of the study is that the QRT+ corrections are found to be qualitatively insignificant and quantitatively small. As the temperature of the bath or the strength of the noise is increased the difference tends to zero with some structure in the noise dependence at low temperature which is an indication of noise induced non-Markovianity. The motional averaging and narrowing caused by the random telegraph noise driving of the TSS was observed in the emission and absorption spectra of the system as function of the noise color with a transition region around $K=0.7$. The decoherence rate of the spectra were found to strongly depend on the noise parameters as they are expected to be directly proportional to the noise frequency $\nu$ and the noise strength $\Omega^2/\nu$ in the slow and fast jumping limits. We have, also, studied the decay rate of the two-time correlation of $\sigma_z$ which can be an indicator of the transport and found that slow noise can enhance transport at low temperature. The reported findings of this study might be beneficial in the open quantum systems when the physical properties corresponding to the two-time correlation functions are of interests and the system-environment interaction might not be described by using a single thermal bath.   

\section{Acknowledgments}
The author would like to acknowledge many useful comments and discussions with Prof. Dr. Resul Eryi\u{g}it.


\begin{thebibliography}{42}%
\makeatletter
\providecommand \@ifxundefined [1]{%
 \@ifx{#1\undefined}
}%
\providecommand \@ifnum [1]{%
 \ifnum #1\expandafter \@firstoftwo
 \else \expandafter \@secondoftwo
 \fi
}%
\providecommand \@ifx [1]{%
 \ifx #1\expandafter \@firstoftwo
 \else \expandafter \@secondoftwo
 \fi
}%
\providecommand \natexlab [1]{#1}%
\providecommand \enquote  [1]{``#1''}%
\providecommand \bibnamefont  [1]{#1}%
\providecommand \bibfnamefont [1]{#1}%
\providecommand \citenamefont [1]{#1}%
\providecommand \href@noop [0]{\@secondoftwo}%
\providecommand \href [0]{\begingroup \@sanitize@url \@href}%
\providecommand \@href[1]{\@@startlink{#1}\@@href}%
\providecommand \@@href[1]{\endgroup#1\@@endlink}%
\providecommand \@sanitize@url [0]{\catcode `\\12\catcode `\$12\catcode
  `\&12\catcode `\#12\catcode `\^12\catcode `\_12\catcode `\%12\relax}%
\providecommand \@@startlink[1]{}%
\providecommand \@@endlink[0]{}%
\providecommand \url  [0]{\begingroup\@sanitize@url \@url }%
\providecommand \@url [1]{\endgroup\@href {#1}{\urlprefix }}%
\providecommand \urlprefix  [0]{URL }%
\providecommand \Eprint [0]{\href }%
\providecommand \doibase [0]{http://dx.doi.org/}%
\providecommand \selectlanguage [0]{\@gobble}%
\providecommand \bibinfo  [0]{\@secondoftwo}%
\providecommand \bibfield  [0]{\@secondoftwo}%
\providecommand \translation [1]{[#1]}%
\providecommand \BibitemOpen [0]{}%
\providecommand \bibitemStop [0]{}%
\providecommand \bibitemNoStop [0]{.\EOS\space}%
\providecommand \EOS [0]{\spacefactor3000\relax}%
\providecommand \BibitemShut  [1]{\csname bibitem#1\endcsname}%
\let\auto@bib@innerbib\@empty
\bibitem [1]{Scully97}%
  \BibitemOpen
  \bibfield  {author} {\bibinfo {author} {\bibfnamefont {M.~O.}\ \bibnamefont
  {Scully}}\ and\ \bibinfo {author} {\bibfnamefont {M.~S.}\ \bibnamefont
  {Zubairy}},\ }\href@noop {} {\emph {\bibinfo {title} {Quantum Optics}}}\
  (\bibinfo  {publisher} {Cambridge University Press},\ \bibinfo {year}
  {1997})\BibitemShut {NoStop}%
\bibitem [2]{Carmichael99}%
  \BibitemOpen
  \bibfield  {author} {\bibinfo {author} {\bibfnamefont {H.~J.}\ \bibnamefont
  {Carmichael}},\ }\href@noop {} {\emph {\bibinfo {title} {Statistical Methods
  in Quantum Optics}}}\ (\bibinfo  {publisher} {Springer Press},\ \bibinfo
  {year} {1999})\BibitemShut {NoStop}%
\bibitem [3]{Gardiner2000}%
  \BibitemOpen
  \bibfield  {author} {\bibinfo {author} {\bibfnamefont {C.~W.}\ \bibnamefont
  {Gardiner}}\ and\ \bibinfo {author} {\bibfnamefont {P.}~\bibnamefont
  {Zoller}},\ }\href@noop {} {\emph {\bibinfo {title} {Quantum Noise}}}\
  (\bibinfo  {publisher} {Springer-Verlag},\ \bibinfo {year}
  {2000})\BibitemShut {NoStop}%
\bibitem [4]{Li2013}%
  \BibitemOpen
  \bibfield  {author} {\bibinfo {author} {\bibfnamefont {J.}\ \bibnamefont
  {Li}}, \bibinfo {author} {\bibfnamefont {M.P.}\ \bibnamefont {Silveri}},
  \bibinfo {author} {\bibfnamefont {K.S.}\ \bibnamefont {Kumar}}, \bibinfo
  {author} {\bibfnamefont {J.-M}\ \bibnamefont {Pirkkalainen}}, \bibinfo
  {author} {\bibfnamefont {A.}~\bibnamefont {Vepsäläinen}}, \bibinfo {author}
  {\bibfnamefont {W.C.}\ \bibnamefont {Chien}}, \bibinfo {author}
  {\bibfnamefont {J.}~\bibnamefont {Tuorila}}, \bibinfo {author} {\bibfnamefont
  {M.A.}\ \bibnamefont {Sillanpää}}, \bibinfo {author} {\bibfnamefont {P.J.}\
  \bibnamefont {Hakonen}}, \bibinfo {author} {\bibfnamefont {E.V.}\
  \bibnamefont {Thuneberg}}, \ and\ \bibinfo {author} {\bibfnamefont {G.S.}\
  \bibnamefont {Paraoanu}},\ }\bibfield  {title} {\enquote {\bibinfo {title}
  {Motional averaging in a superconducting qubit},}\ }\href@noop {} {\bibfield
  {journal} {\bibinfo  {journal} {Nat. commun.}\ }\textbf {\bibinfo {volume}
  {4}},\ \bibinfo {pages} {1420} (\bibinfo {year} {2013})}\BibitemShut
  {NoStop}%
\bibitem [5]{Li2018}%
  \BibitemOpen
  \bibfield  {author} {\bibinfo {author} {\bibfnamefont {L.}~\bibnamefont
  {Li}}, \bibinfo {author} {\bibfnamefont {M.~J.W.}\ \bibnamefont {Hall}},
  \ and\ \bibinfo {author} {\bibfnamefont {H.~M.}\ \bibnamefont
  {Wisemanv}},\ }\bibfield  {title} {\enquote {\bibinfo {title} {Concepts of
  quantum non-markovianity: A hierarchy},}\ }\href@noop {} {\bibfield
  {journal} {\bibinfo  {journal} {Phy. Rep.}\ }\textbf {\bibinfo {volume}
  {759}},\ \bibinfo {pages} {1--51} (\bibinfo {year} {2018})}\BibitemShut
  {NoStop}%
\bibitem [6]{deVega2017}%
  \BibitemOpen
  \bibfield  {author} {\bibinfo {author} {\bibfnamefont {I.}\ \bibnamefont
  {de~Vega}}\ and\ \bibinfo {author} {\bibfnamefont {D.}\ \bibnamefont
  {Alonso}},\ }\bibfield  {title} {\enquote {\bibinfo {title} {Dynamics of
  non-markovian open quantum systems},}\ }\href@noop {} {\bibfield  {journal}
  {\bibinfo  {journal} {Rev. Mod. Phys.}\ }\textbf {\bibinfo {volume} {89}},\
  \bibinfo {pages} {015001} (\bibinfo {year} {2017})}\BibitemShut {NoStop}%
\bibitem [7]{Bertelot2006}%
  \BibitemOpen
  \bibfield  {author} {\bibinfo {author} {\bibfnamefont {A.}~\bibnamefont
  {Berthelot}}, \bibinfo {author} {\bibfnamefont {I.}~\bibnamefont {Favero}},
  \bibinfo {author} {\bibfnamefont {G.}~\bibnamefont {Cassabois}}, \bibinfo
  {author} {\bibfnamefont {C.}~\bibnamefont {Voisin}}, \bibinfo {author}
  {\bibfnamefont {C.}~\bibnamefont {Delalande}}, \bibinfo {author}
  {\bibfnamefont {Ph.}\ \bibnamefont {Roussignol}}, \bibinfo {author}
  {\bibfnamefont {R.}~\bibnamefont {Ferreira}}, \ and\ \bibinfo {author}
  {\bibfnamefont {J.~M.}\ \bibnamefont {Gérard}},\ }\bibfield  {title}
  {\enquote {\bibinfo {title} {Unconventional motional narrowing in the optical
  spectrum of a semiconductor quantum dot},}\ }\href@noop {} {\bibfield
  {journal} {\bibinfo  {journal} {Nat. Phys.}\ }\textbf {\bibinfo {volume}
  {2}},\ \bibinfo {pages} {759--764} (\bibinfo {year} {2006})}\BibitemShut
  {NoStop}%
\bibitem [8]{Winger2009}%
  \BibitemOpen
  \bibfield  {author} {\bibinfo {author} {\bibfnamefont {M.}\ \bibnamefont
  {Winger}}, \bibinfo {author} {\bibfnamefont {T.}\ \bibnamefont {Volz}},
  \bibinfo {author} {\bibfnamefont {G.}\ \bibnamefont {Tarel}}, \bibinfo
  {author} {\bibfnamefont {S.}\ \bibnamefont {Portolan}}, \bibinfo
  {author} {\bibfnamefont {A.}\ \bibnamefont {Badolato}}, \bibinfo
  {author} {\bibfnamefont {K.~J.}\ \bibnamefont {Hennessy}}, \bibinfo
  {author} {\bibfnamefont {E.~L.}\ \bibnamefont {Hu}}, \bibinfo {author}
  {\bibfnamefont {A.}\ \bibnamefont {Beveratos}}, \bibinfo {author}
  {\bibfnamefont {J.}\ \bibnamefont {Finley}}, \bibinfo {author}
  {\bibfnamefont {V.}\ \bibnamefont {Savona}}, \ and\ \bibinfo {author}
  {\bibfnamefont {A.}\ \bibnamefont
  {Imamo\ifmmode~\breve{g}\else \u{g}\fi{}lu}},\ }\bibfield  {title} {\enquote
  {\bibinfo {title} {Explanation of photon correlations in the
  far-off-resonance optical emission from a quantum-dot--cavity system},}\
  }\href@noop {} {\bibfield  {journal} {\bibinfo  {journal} {Phys. Rev. Lett.}\
  }\textbf {\bibinfo {volume} {103}},\ \bibinfo {pages} {207403} (\bibinfo
  {year} {2009})}\BibitemShut {NoStop}%
\bibitem [9]{Ulrich2011}%
  \BibitemOpen
  \bibfield  {author} {\bibinfo {author} {\bibfnamefont {S.~M.}\ \bibnamefont
  {Ulrich}}, \bibinfo {author} {\bibfnamefont {S.}~\bibnamefont {Ates}},
  \bibinfo {author} {\bibfnamefont {S.}~\bibnamefont {Reitzenstein}}, \bibinfo
  {author} {\bibfnamefont {A.}~\bibnamefont {L\"offler}}, \bibinfo {author}
  {\bibfnamefont {A.}~\bibnamefont {Forchel}}, \ and\ \bibinfo {author}
  {\bibfnamefont {P.}~\bibnamefont {Michler}},\ }\bibfield  {title} {\enquote
  {\bibinfo {title} {Dephasing of triplet-sideband optical emission of a
  resonantly driven $\mathrm{InAs}/\mathrm{GaAs}$ quantum dot inside a
  microcavity},}\ }\href@noop {} {\bibfield  {journal} {\bibinfo  {journal}
  {Phys. Rev. Lett.}\ }\textbf {\bibinfo {volume} {106}},\ \bibinfo {pages}
  {247402} (\bibinfo {year} {2011})}\BibitemShut {NoStop}%
\bibitem [10]{Jorgensen2020}%
  \BibitemOpen
  \bibfield  {author} {\bibinfo {author} {\bibfnamefont {Mathias~R.}\
  \bibnamefont {J\o{}rgensen}}\ and\ \bibinfo {author} {\bibfnamefont
  {Felix~A.}\ \bibnamefont {Pollock}},\ }\bibfield  {title} {\enquote {\bibinfo
  {title} {A discrete memory-kernel for multi-time correlations in
  non-markovian quantum processes},}\ }\href@noop {} {\bibfield  {journal}
  {\bibinfo  {journal} {Phys. Rev. A}\ }\textbf {\bibinfo {volume} {102}},\
  \bibinfo {pages} {052206} (\bibinfo {year} {2020})}\BibitemShut {NoStop}%
\bibitem[11]{Brixner2005}%
  \BibitemOpen
  \bibfield  {author} {\bibinfo {author} {\bibfnamefont {T.}\ \bibnamefont
  {Brixner}}, \bibinfo {author} {\bibfnamefont {J.}\ \bibnamefont {Stenger}},
  \bibinfo {author} {\bibfnamefont {H.~M.}\ \bibnamefont {Vaswani}},
  \bibinfo {author} {\bibfnamefont {M.}\ \bibnamefont {Cho}}, \bibinfo
  {author} {\bibfnamefont {R.~E.}\ \bibnamefont {Blankenship}}, \ and\
  \bibinfo {author} {\bibfnamefont {G.~R.}\ \bibnamefont {Fleming}},\
  }\bibfield  {title} {\enquote {\bibinfo {title} {Two-dimensional spectroscopy
  of electronic couplings in photosynthesis},}\ }\href@noop {} {\bibfield
  {journal} {\bibinfo  {journal} {Nature}\ }\textbf {\bibinfo {volume} {434}},\
  \bibinfo {pages} {625–628} (\bibinfo {year} {2005})}\BibitemShut {NoStop}%
\bibitem [12]{Onsager31a}%
  \BibitemOpen
  \bibfield  {author} {\bibinfo {author} {\bibfnamefont {L.}\ \bibnamefont
  {Onsager}},\ }\bibfield  {title} {\enquote {\bibinfo {title} {Reciprocal
  relations in irreversible processes. i.}}\ }\href@noop {} {\bibfield
  {journal} {\bibinfo  {journal} {Phys. Rev.}\ }\textbf {\bibinfo {volume}
  {37}},\ \bibinfo {pages} {405} (\bibinfo {year} {1931})}\BibitemShut
  {NoStop}%
\bibitem [13]{Lax63}%
  \BibitemOpen
  \bibfield  {author} {\bibinfo {author} {\bibfnamefont {M.}\ \bibnamefont
  {Lax}},\ }\bibfield  {title} {\enquote {\bibinfo {title} {Formal theory of
  quantum fluctuations from a driven state},}\ }\href@noop {} {\bibfield
  {journal} {\bibinfo  {journal} {Phys. Rev.}\ }\textbf {\bibinfo {volume}
  {129}},\ \bibinfo {pages} {2342--2348} (\bibinfo {year} {1963})}\BibitemShut
  {NoStop}%
\bibitem[14]{Alonso2005}%
  \BibitemOpen
  \bibfield  {author} {\bibinfo {author} {\bibfnamefont {D.}\ \bibnamefont
  {Alonso}}\ and\ \bibinfo {author} {\bibfnamefont {I.}\ \bibnamefont
  {de~Vega}},\ }\bibfield  {title} {\enquote {\bibinfo {title} {Multiple-time
  correlation functions for non-markovian interaction: Beyond the quantum
  regression theorem},}\ }\href@noop {} {\bibfield  {journal} {\bibinfo
  {journal} {Phys. Rev. Lett.}\ }\textbf {\bibinfo {volume} {94}},\ \bibinfo
  {pages} {200403} (\bibinfo {year} {2005})}\BibitemShut {NoStop}%
\bibitem[15]{deVega2006}%
  \BibitemOpen
  \bibfield  {author} {\bibinfo {author} {\bibfnamefont {I.}\ \bibnamefont
  {de~Vega}}\ and\ \bibinfo {author} {\bibfnamefont {D.}\ \bibnamefont
  {Alonso}},\ }\bibfield  {title} {\enquote {\bibinfo {title} {Non-markovian
  reduced propagator, multiple-time correlation functions, and master equations
  with general initial conditions in the weak-coupling limit},}\ }\href@noop {}
  {\bibfield  {journal} {\bibinfo  {journal} {Phys. Rev. A}\ }\textbf {\bibinfo
  {volume} {73}},\ \bibinfo {pages} {022102} (\bibinfo {year}
  {2006})}\BibitemShut {NoStop}%
\bibitem[16]{Alonso2007}%
  \BibitemOpen
  \bibfield  {author} {\bibinfo {author} {\bibfnamefont {D.}~\bibnamefont
  {Alonso}}\ and\ \bibinfo {author} {\bibfnamefont {I.}~\bibnamefont
  {de~Vega}},\ }\bibfield  {title} {\enquote {\bibinfo {title} {Hierarchy of
  equations of multiple-time correlation functions},}\ }\href@noop {}
  {\bibfield  {journal} {\bibinfo  {journal} {Phys. Rev. A}\ }\textbf {\bibinfo
  {volume} {75}},\ \bibinfo {pages} {052108} (\bibinfo {year}
  {2007})}\BibitemShut {NoStop}%
\bibitem[17]{Budini2008}%
  \BibitemOpen
  \bibfield  {author} {\bibinfo {author} {\bibfnamefont {A.~A.}\
  \bibnamefont {Budini}},\ }\bibfield  {title} {\enquote {\bibinfo {title}
  {Operator correlations and quantum regression theorem in non-markovian
  lindblad rate equations},}\ }\href@noop {} {\bibfield  {journal} {\bibinfo
  {journal} {J. Stat. Phys.}\ }\textbf {\bibinfo {volume}
  {131}},\ \bibinfo {pages} {51--78} (\bibinfo {year} {2008})}\BibitemShut
  {NoStop}%
\bibitem[18]{goan11}%
  \BibitemOpen
  \bibfield  {author} {\bibinfo {author} {\bibfnamefont {H.-S.}\ \bibnamefont
  {Goan}}, \bibinfo {author} {\bibfnamefont {P.-W.}\ \bibnamefont {Chen}}, \
  and\ \bibinfo {author} {\bibfnamefont {C.-C.}\ \bibnamefont {Jian}},\
  }\bibfield  {title} {\enquote {\bibinfo {title} {Non-markovian
  finite-temperature two-time correlation functions of system operators: Beyond
  the quantum regression theorem},}\ }\href@noop {} {\bibfield  {journal}
  {\bibinfo  {journal} {J. Chem. Phys.}\ }\textbf {\bibinfo {volume} {134}},\
  \bibinfo {pages} {124112} (\bibinfo {year} {2011})}\BibitemShut {NoStop}%
\bibitem[19]{chen2011}%
  \BibitemOpen
  \bibfield  {author} {\bibinfo {author} {\bibfnamefont {P.-W.}\ \bibnamefont
  {Chen}}, \bibinfo {author} {\bibfnamefont {C.-C.}\ \bibnamefont {Jian}},
  \ and\ \bibinfo {author} {\bibfnamefont {H.-S.}\ \bibnamefont {Goan}},\
  }\bibfield  {title} {\enquote {\bibinfo {title} {Non-markovian dynamics of a
  nanomechanical resonator measured by a quantum point contact},}\ }\href@noop
  {} {\bibfield  {journal} {\bibinfo  {journal} {Phys. Rev. B}\ }\textbf
  {\bibinfo {volume} {83}},\ \bibinfo {pages} {115439} (\bibinfo {year}
  {2011})}\BibitemShut {NoStop}%
\bibitem[20]{Goan2010}%
  \BibitemOpen
  \bibfield  {author} {\bibinfo {author} {\bibfnamefont {H.-S.}\
  \bibnamefont {Goan}}, \bibinfo {author} {\bibfnamefont {C.-C.}\
  \bibnamefont {Jian}}, \ and\ \bibinfo {author} {\bibfnamefont {P.-W.}\
  \bibnamefont {Chen}},\ }\bibfield  {title} {\enquote {\bibinfo {title}
  {Non-markovian finite-temperature two-time correlation functions of system
  operators of a pure-dephasing model},}\ }\href@noop {} {\bibfield  {journal}
  {\bibinfo  {journal} {Phys. Rev. A}\ }\textbf {\bibinfo {volume} {82}},\
  \bibinfo {pages} {012111} (\bibinfo {year} {2010})}\BibitemShut {NoStop}%
\bibitem[21]{dara2016}%
  \BibitemOpen
  \bibfield  {author} {\bibinfo {author} {\bibfnamefont {D.~P.~S.}\
  \bibnamefont {McCutcheon}},\ }\bibfield  {title} {\enquote {\bibinfo {title}
  {Optical signatures of non-markovian behavior in open quantum systems},}\
  }\href@noop {} {\bibfield  {journal} {\bibinfo  {journal} {Phys. Rev. A}\
  }\textbf {\bibinfo {volume} {93}},\ \bibinfo {pages} {022119} (\bibinfo
  {year} {2016})}\BibitemShut {NoStop}%
\bibitem[22]{deVega2008}%
  \BibitemOpen
  \bibfield  {author} {\bibinfo {author} {\bibfnamefont {I.}\ \bibnamefont
  {de~Vega}}\ and\ \bibinfo {author} {\bibfnamefont {D.}\ \bibnamefont
  {Alonso}},\ }\bibfield  {title} {\enquote {\bibinfo {title} {Emission spectra
  of atoms with non-markovian interaction: Fluorescence in a photonic
  crystal},}\ }\href@noop {} {\bibfield  {journal} {\bibinfo  {journal} {Phys.
  Rev. A}\ }\textbf {\bibinfo {volume} {77}},\ \bibinfo {pages} {043836}
  (\bibinfo {year} {2008})}\BibitemShut {NoStop}%
\bibitem[23]{Jin2016}%
  \BibitemOpen
  \bibfield  {author} {\bibinfo {author} {\bibfnamefont {J.}\
  \bibnamefont {Jin}}, \bibinfo {author} {\bibfnamefont {C.}\
  \bibnamefont {Karlewski}}, \ and\ \bibinfo {author} {\bibfnamefont {M.}\
  \bibnamefont {Marthaler}},\ }\bibfield  {title} {\enquote {\bibinfo {title}
  {Non-markovian correlation functions for open quantum systems},}\ }\href@noop
  {} {\bibfield  {journal} {\bibinfo  {journal} {New J. Phys.}\ }\textbf
  {\bibinfo {volume} {18}},\ \bibinfo {pages} {083038} (\bibinfo {year}
  {2016})}\BibitemShut {NoStop}%
\bibitem[24]{Ford1996}%
  \BibitemOpen
  \bibfield  {author} {\bibinfo {author} {\bibfnamefont {G.~W.}\ \bibnamefont
  {Ford}}\ and\ \bibinfo {author} {\bibfnamefont {R.~F.}\ \bibnamefont
  {O'Connell}},\ }\bibfield  {title} {\enquote {\bibinfo {title} {There is no
  quantum regression theorem},}\ }\href@noop {} {\bibfield  {journal} {\bibinfo
   {journal} {Phys. Rev. Lett.}\ }\textbf {\bibinfo {volume} {77}},\ \bibinfo
  {pages} {798--801} (\bibinfo {year} {1996})}\BibitemShut {NoStop}%
\bibitem[25]{Ford1999}%
  \BibitemOpen
  \bibfield  {author} {\bibinfo {author} {\bibfnamefont {G.~W.}\ \bibnamefont
  {Ford}}\ and\ \bibinfo {author} {\bibfnamefont {R.~F.}\ \bibnamefont
  {O'Connell}},\ }\bibfield  {title} {\enquote {\bibinfo {title} {Calculation
  of correlation functions in the weak coupling approximation},}\ }\href@noop
  {} {\bibfield  {journal} {\bibinfo  {journal} {Ann. Phys.}\ }\textbf
  {\bibinfo {volume} {276}},\ \bibinfo {pages} {144--151} (\bibinfo {year}
  {1999})}\BibitemShut {NoStop}%
\bibitem[26]{Ford2000}%
  \BibitemOpen
  \bibfield  {author} {\bibinfo {author} {\bibfnamefont {G.~W.}\ \bibnamefont
  {Ford}}\ and\ \bibinfo {author} {\bibfnamefont {R.~F.}\ \bibnamefont
  {O'Connell}},\ }\bibfield  {title} {\enquote {\bibinfo {title} {Driven
  systems and the lax formula},}\ }\href@noop {} {\bibfield  {journal}
  {\bibinfo  {journal} {Opt. Commun.}\ }\textbf {\bibinfo {volume}
  {179}},\ \bibinfo {pages} {451--461} (\bibinfo {year} {2000})}\BibitemShut
  {NoStop}%
\bibitem[27]{Blanter2000}%
  \BibitemOpen
  \bibfield  {author} {\bibinfo {author} {\bibfnamefont {Ya.M.}\ \bibnamefont
  {Blanter}}\ and\ \bibinfo {author} {\bibfnamefont {M.}~\bibnamefont
  {Büttiker}},\ }\bibfield  {title} {\enquote {\bibinfo {title} {Shot noise in
  mesoscopic conductors},}\ }\href@noop {} {\bibfield  {journal} {\bibinfo
  {journal} {Phys. Rep.}\ }\textbf {\bibinfo {volume} {336}},\ \bibinfo
  {pages} {1--166} (\bibinfo {year} {2000})}\BibitemShut {NoStop}%
\bibitem[28]{Cosacchi2018}%
  \BibitemOpen
  \bibfield  {author} {\bibinfo {author} {\bibfnamefont {M.}~\bibnamefont
  {Cosacchi}}, \bibinfo {author} {\bibfnamefont {M.}~\bibnamefont {Cygorek}},
  \bibinfo {author} {\bibfnamefont {F.}~\bibnamefont {Ungar}}, \bibinfo
  {author} {\bibfnamefont {A.~M.}\ \bibnamefont {Barth}}, \bibinfo {author}
  {\bibfnamefont {A.}~\bibnamefont {Vagov}}, \ and\ \bibinfo {author}
  {\bibfnamefont {V.~M.}\ \bibnamefont {Axt}},\ }\bibfield  {title} {\enquote
  {\bibinfo {title} {Path-integral approach for nonequilibrium multitime
  correlation functions of open quantum systems coupled to markovian and
  non-markovian environments},}\ }\href@noop {} {\bibfield  {journal} {\bibinfo
   {journal} {Phys. Rev. B}\ }\textbf {\bibinfo {volume} {98}},\ \bibinfo
  {pages} {125302} (\bibinfo {year} {2018})}\BibitemShut {NoStop}%
\bibitem[29]{Shao2002}%
  \BibitemOpen
  \bibfield  {author} {\bibinfo {author} {\bibfnamefont {J.}~\bibnamefont
  {Shao}}\ and\ \bibinfo {author} {\bibfnamefont {N.}~\bibnamefont {Makri}},\
  }\bibfield  {title} {\enquote {\bibinfo {title} {Iterative path integral
  formulation of equilibrium correlation functions for quantum dissipative
  systems},}\ }\href@noop {} {\bibfield  {journal} {\bibinfo  {journal} {J.
  Chem. Phys.}\ }\textbf {\bibinfo {volume} {116}},\ \bibinfo {pages} {507}
  (\bibinfo {year} {2002})}\BibitemShut {NoStop}%
\bibitem[30]{Pollock2018}%
  \BibitemOpen
  \bibfield  {author} {\bibinfo {author} {\bibfnamefont {F.}~\bibnamefont
  {Pollock}}, \bibinfo {author} {\bibfnamefont {C.~A.}\ \bibnamefont
  {Rodriguez-Rosario}}, \bibinfo {author} {\bibfnamefont {T.}~\bibnamefont
  {Frauenheim}}, \bibinfo {author} {\bibfnamefont {M.}~\bibnamefont
  {Paternostro}}, \ and\ \bibinfo {author} {\bibfnamefont {K.}~\bibnamefont
  {Modi}},\ }\bibfield  {title} {\enquote {\bibinfo {title} {Non-markovian
  quantum processes: Complete framework and efficient characterization},}\
  }\href@noop {} {\bibfield  {journal} {\bibinfo  {journal} {Phys. Rev. A}\ }\textbf {\bibinfo {volume} {97}},\ \bibinfo {pages} {012127} (\bibinfo
  {year} {2018})}\BibitemShut {NoStop}%
\bibitem[31]{Jorgensen2019}%
  \BibitemOpen
  \bibfield  {author} {\bibinfo {author} {\bibfnamefont {M.~R.}\
  \bibnamefont {J\o{}rgensen}}\ and\ \bibinfo {author} {\bibfnamefont
  {F.~A.}\ \bibnamefont {Pollock}},\ }\bibfield  {title} {\enquote {\bibinfo
  {title} {Exploiting the causal tensor network structure of quantum processes
  to efficiently simulate non-markovian path integrals},}\ }\href@noop {}
  {\bibfield  {journal} {\bibinfo  {journal} {Phys. Rev. Lett.}\ }\textbf
  {\bibinfo {volume} {123}},\ \bibinfo {pages} {240602} (\bibinfo {year}
  {2019})}\BibitemShut {NoStop}%
\bibitem[32]{Guarnieri2014}%
  \BibitemOpen
  \bibfield  {author} {\bibinfo {author} {\bibfnamefont {G.}\ \bibnamefont
  {Guarnieri}}, \bibinfo {author} {\bibfnamefont {A.}\ \bibnamefont
  {Smirne}}, \ and\ \bibinfo {author} {\bibfnamefont {B.}\ \bibnamefont
  {Vacchini}},\ }\bibfield  {title} {\enquote {\bibinfo {title} {Quantum
  regression theorem and non-markovianity of quantum dynamics},}\ }\href@noop
  {} {\bibfield  {journal} {\bibinfo  {journal} {Phys. Rev. A}\ }\textbf
  {\bibinfo {volume} {90}},\ \bibinfo {pages} {022110} (\bibinfo {year}
  {2014})}\BibitemShut {NoStop}%
\bibitem[33]{Chen2014}%
  \BibitemOpen
  \bibfield  {author} {\bibinfo {author} {\bibfnamefont {P.-W.}\ \bibnamefont
  {Chen}}\ and\ \bibinfo {author} {\bibfnamefont {Md.~M.}\ \bibnamefont
  {Ali}},\ }\bibfield  {title} {\enquote {\bibinfo {title} {Investigating
  leggett-garg inequality for a two level system under decoherence in a
  non-markovian dephasing environment},}\ }\href@noop {} {\bibfield  {journal}
  {\bibinfo  {journal} {Sci. Rep.}\ }\textbf {\bibinfo {volume} {4}},\
  \bibinfo {pages} {6165} (\bibinfo {year} {2014})}\BibitemShut {NoStop}%
\bibitem[34]{Ali2015}%
  \BibitemOpen
  \bibfield  {author} {\bibinfo {author} {\bibfnamefont {Md.~M.}\
  \bibnamefont {Ali}}, \bibinfo {author} {\bibfnamefont {P.-Y.}\
  \bibnamefont {L.}}, \bibinfo {author} {\bibfnamefont {M. W.-Y.}\
  \bibnamefont {Tu}}, \ and\ \bibinfo {author} {\bibfnamefont {W.-M.}\
  \bibnamefont {Zhang}},\ }\bibfield  {title} {\enquote {\bibinfo {title}
  {Non-markovianity measure using two-time correlation functions},}\
  }\href@noop {} {\bibfield  {journal} {\bibinfo  {journal} {Phys. Rev. A}\
  }\textbf {\bibinfo {volume} {92}},\ \bibinfo {pages} {062306} (\bibinfo
  {year} {2015})}\BibitemShut {NoStop}%
  
  
  \bibitem[35]{Kurt2020}%
  \BibitemOpen
  \bibfield  {author} {\bibinfo {author} {\bibfnamefont {A.}\ \bibnamefont
  {Kurt}},\ }\bibfield
   {title} {\enquote {\bibinfo {title} {Non-Markovian Corrections to Quantum Regression Theorem for the Strong Coupling Spin-Boson Model},}\ }\href@noop {} {\bibfield  {journal} {\bibinfo
  {journal} {SAUJS}\ }\textbf {\bibinfo {volume} {24}},\ \bibinfo
  {pages} {596--604} (\bibinfo {year} {2020})}\BibitemShut {NoStop}%
  
  \bibitem[36]{Jang2018}%
  \BibitemOpen
  \bibfield  {author} {\bibinfo {author} {\bibfnamefont {S.~J.}\ \bibnamefont
  {Jang}}, \bibinfo {author} {\bibfnamefont {M.}\ \bibnamefont { Benedetta}},\ }\bibfield
   {title} {\enquote {\bibinfo {title} {Delocalized excitons in natural light-harvesting complexes},}\ }\href@noop {} {\bibfield  {journal} {\bibinfo
  {journal} {Rev. Mod. Phys.}\ }\textbf {\bibinfo {volume} {90}},\ \bibinfo
  {pages} {035003} (\bibinfo {year} {2018})}\BibitemShut {NoStop}%
  
  
\bibitem[37]{Garg1985}%
  \BibitemOpen
  \bibfield  {author} {\bibinfo {author} {\bibfnamefont {A.}~\bibnamefont
  {Garg}}, \bibinfo {author} {\bibfnamefont {J.~N.}\ \bibnamefont {Onuchic}}, \
  and\ \bibinfo {author} {\bibfnamefont {V.~J.}\ \bibnamefont {Ambegaokar}},\
  }\bibfield  {title} {\enquote {\bibinfo {title} {Effect of friction on
  electron transfer in biomolecules},}\ }\href@noop {} {\bibfield  {journal}
  {\bibinfo  {journal} {J. Chem. Phys.}\ }\textbf {\bibinfo {volume} {83}},\
  \bibinfo {pages} {4491--4503} (\bibinfo {year} {1985})}\BibitemShut {NoStop}%
\bibitem[38]{Bourret73}%
  \BibitemOpen
  \bibfield  {author} {\bibinfo {author} {\bibfnamefont {R.~C.}\ \bibnamefont
  {Bourret}}, \bibinfo {author} {\bibfnamefont {U.}~\bibnamefont {Frisch}}, \
  and\ \bibinfo {author} {\bibfnamefont {A.}~\bibnamefont {Pouquet}},\
  }\bibfield  {title} {\enquote {\bibinfo {title} {Browian motion of harmonic
  oscillator with stochastic frequency},}\ }\href@noop {} {\bibfield  {journal}
  {\bibinfo  {journal} {Physica}\ }\textbf {\bibinfo {volume} {65}},\ \bibinfo
  {pages} {303--320} (\bibinfo {year} {1973})}\BibitemShut {NoStop}%
\bibitem[39]{Shapiro78}%
  \BibitemOpen
  \bibfield  {author} {\bibinfo {author} {\bibfnamefont {V.~E.}\ \bibnamefont
  {Shapiro}}\ and\ \bibinfo {author} {\bibfnamefont {V.~M.}\ \bibnamefont
  {Loginov}},\ }\bibfield  {title} {\enquote {\bibinfo {title} {"formulae of
  differentiation" and their use for solving stochastic equations},}\
  }\href@noop {} {\bibfield  {journal} {\bibinfo  {journal} {Physica A}\
  }\textbf {\bibinfo {volume} {91}},\ \bibinfo {pages} {563} (\bibinfo {year}
  {1978})}\BibitemShut {NoStop}%
\bibitem[40]{Magazzu2017}%
  \BibitemOpen
  \bibfield  {author} {\bibinfo {author} {\bibfnamefont {L.}\ \bibnamefont
  {Magazz\`u}}, \bibinfo {author} {\bibfnamefont {P.}\ \bibnamefont
  {H\"anggi}}, \bibinfo {author} {\bibfnamefont {B.}\ \bibnamefont
  {Spagnolo}}, \ and\ \bibinfo {author} {\bibfnamefont {D.}\ \bibnamefont
  {Valenti}},\ }\bibfield  {title} {\enquote {\bibinfo {title} {Quantum
  resonant activation},}\ }\href@noop {} {\bibfield  {journal} {\bibinfo
  {journal} {Phys. Rev. E}\ }\textbf {\bibinfo {volume} {95}},\ \bibinfo
  {pages} {042104} (\bibinfo {year} {2017})}\BibitemShut {NoStop}%
\bibitem[41]{Goychuk95}%
  \BibitemOpen
  \bibfield  {author} {\bibinfo {author} {\bibfnamefont {I.~A.}\ \bibnamefont
  {Goychuk}}, \bibinfo {author} {\bibfnamefont {E.~G.}\ \bibnamefont {Petrov}},
  \ and\ \bibinfo {author} {\bibfnamefont {V.}~\bibnamefont {May}},\ }\bibfield
   {title} {\enquote {\bibinfo {title} {Dynamics of the dissipative two-level
  system driven by external telegraph noise},}\ }\href@noop {} {\bibfield
  {journal} {\bibinfo  {journal} {Phys. Rev. E}\ }\textbf {\bibinfo {volume}
  {52}},\ \bibinfo {pages} {2392--2400} (\bibinfo {year} {1995})}\BibitemShut
  {NoStop}%
\bibitem[42]{Berry95}%
  \BibitemOpen
  \bibfield  {author} {\bibinfo {author} {\bibfnamefont {M.}\ \bibnamefont
  {Berry}},\ }\bibfield  {title} {\enquote {\bibinfo {title} {Two-state quantum
  asymptotics},}\ }\href@noop {} {\bibfield  {journal} {\bibinfo  {journal}
  {Ann. N. Y. Acad. Sci.}\ }\textbf {\bibinfo {volume} {755}},\ \bibinfo
  {pages} {303} (\bibinfo {year} {1995})}\BibitemShut {NoStop}%
\bibitem[43]{Rebentrost2009}%
  \BibitemOpen
  \bibfield  {author} {\bibinfo {author} {\bibfnamefont {P.}\ \bibnamefont
  {Rebentrost}}, \bibinfo {author} {\bibfnamefont {M.}\ \bibnamefont
  {Mohseni}}, \bibinfo {author} {\bibfnamefont {I.}\ \bibnamefont {Kassal}},
  \bibinfo {author} {\bibfnamefont {S.}\ \bibnamefont {Lloyd}}, \ and\
  \bibinfo {author} {\bibfnamefont {A.}\ \bibnamefont {Aspuru-Guzik}},\
  }\bibfield  {title} {\enquote {\bibinfo {title} {Environment-assisted quantum
  transport},}\ }\href@noop {} {\bibfield  {journal} {\bibinfo  {journal} {New
  J. Phys.}\ }\textbf {\bibinfo {volume} {11}},\ \bibinfo {pages} {033003}
  (\bibinfo {year} {2009})}\BibitemShut {NoStop}%
  
\bibitem[44]{Wang2005}%
  \BibitemOpen
  \bibfield  {author} {\bibinfo {author} {\bibfnamefont {X.~R.}\ \bibnamefont
  {Wang}}, \bibinfo {author} {\bibfnamefont {Y.~S.}\ \bibnamefont {Zheng}}, \
  and\ \bibinfo {author} {\bibfnamefont {Sun}\ \bibnamefont {Yin}},\ }\bibfield
   {title} {\enquote {\bibinfo {title} {Spin relaxation and decoherence of
  two-level systems},}\ }\href@noop {} {\bibfield  {journal} {\bibinfo
  {journal} {Phys. Rev. B}\ }\textbf {\bibinfo {volume} {72}},\ \bibinfo
  {pages} {121303} (\bibinfo {year} {2005})}\BibitemShut {NoStop}%
   
\end{thebibliography}

%

\end{document}